\documentclass[a4paper,fleqn,usenatbib]{mnras}

\usepackage[T1]{fontenc}
\usepackage{ae,aecompl}

\usepackage{graphicx}   % Including figure files
\usepackage{amsmath}    % Advanced maths commands
\usepackage{amssymb}    % Extra maths symbols

\title{The distance sum rule from strong lensing systems and quasars - test of cosmic curvature and beyond}

\author[Jing-Zhao Qi et al.]
{Jing-Zhao Qi$^{1,2}$, Shuo Cao$^{2}$,\thanks{E-mail:
caoshuo@bnu.edu.cn} Sixuan Zhang$^{2}$, Marek
Biesiada$^{2,3}$,\thanks{E-mail: marek.biesiada@us.edu.pl} Yan
Wu$^{2}$
\newauthor and Zong-Hong Zhu$^{2}$\\
$^{1}$ Department of Physics, College of Sciences, Northeastern University, Shenyang 110819, China;\\
$^{2}$ Department of Astronomy, Beijing Normal University, Beijing 100875, China;\\
$^{3}$ Department of Astrophysics and Cosmology, Institute of
Physics, University of Silesia, 75 Pu{\l}ku Piechoty 1, 41-500
Chorz{\'o}w, Poland }

\date{Accepted XXX. Received YYY; in original form ZZZ}

\pubyear{2018}

\begin{document}
\label{firstpage}
\pagerange{\pageref{firstpage}--\pageref{lastpage}}
\maketitle

\begin{abstract}

Testing the distance-sum-rule in strong lensing systems provides an
interesting method to determine the curvature parameter $\Omega_k$
using more local objects. In this paper, we apply this method to a
quite recent data set of strong lensing systems in combination with
intermediate-luminosity quasars calibrated as standard rulers. In
the framework of three types of lens models extensively used in
strong lensing studies (SIS model, power-law spherical model, and
extended power-law lens model), we show that the assumed lens model
has a considerable impact on the cosmic curvature constraint, which
is found to be compatible or marginally compatible with the flat
case (depending on the lens model adopted). Analysis of low,
intermediate and high-mass sub-samples defined according to the lens
velocity dispersion demonstrates that, although it is not reasonable
to characterize all lenses with a uniform model, such division has
little impact on cosmic curvature inferred. Finally, thinking about
future when massive surveys will provide their yields, we simulated
a mock catalog of strong lensing systems expected to be seen by the
LSST, together with a realistic catalog of quasars. We found that
with about 16000 such systems, combined with the distance
information provided by 500 compact milliarcsecond radio sources
seen in future radio astronomical surveys, one would be able to
constrain the cosmic curvature with an accuracy of $\Delta
\Omega_k\simeq 10^{-3}$, which is comparable to the precision of
\textit{Planck} 2015 results.

\end{abstract}


\section{Introduction} \label{introduction}

Since the beginning of modern cosmology, the curvature parameter
$\Omega_k$ of the Universe has been one of the most important
cosmological parameters one attempted to measure. Its value, or even
its sign, bears important information on the matter-energy budget in
the Universe and sheds light on its ultimate fate. Moreover, it
clarifies which one of three possible classes of
Friedman-Lema\^{\i}tre-Robertson-Walker (FLRW) metric correctly describes our
Universe and is adequate for cosmological inference. New, precise
measurements of anisotropies in the cosmic microwave background
radiation (CMBR) convinced us that the Universe we live in is
spatially flat as it was expected by the supporters of inflationary
models. The latest, most precise constraint, $|\Omega_k|<0.005$,
comes from the {\it Planck} satellite's CMBR data
\citep{ade2016planck}. Such constraints are possible since the
acoustic horizon scale, corresponding to the distance that sound
waves could have traveled between the birth of the universe and the
time when the CMBR was emitted (the surface of last scattering) can
be calculated theoretically. So it can be used as a standard ruler
imprinted in the CMBR anisotropy patterns. Its apparent size depends
on whether the Universe is flat, negatively or positively curved,
which underlies the measurement of curvature parameter from CMBR
anisotropies. The length of this standard ruler is somewhat
degenerate with quantities such as the matter density or dark energy
equation of state. Therefore, cosmic curvature inferred from CMBR is
model-dependent to some extent \citep{ade2016planck}.

New dimension to the significance of cosmic curvature measurements
has been given by the works of
\citet{clarkson2008general,rasanen2015new}. They realized that
measurements of the cosmic curvature using extragalactic objects
(standard rulers and standard candles detected in future all-sky
surveys) could serve as a test of the Copernican Principle i.e. the
validity of the FLRW metric. Moreover, \citet{Bolejko17a,Bolejko17b}
examined the emergence of negative spatial curvature within the so
called silent universe in which the back-reaction of inhomogeneities
(coming from the structure formation) was taken into account. This
effect could potentially be detected using closer objects instead of
CMBR. Measurements of $\Omega_k$ according to the idea of
\citet{clarkson2008general} have been performed using the current
observational data comprising supernovae Ia (SN Ia) (standard
candles) and Hubble parameters $H(z)$ inferred from cosmic
chronometers
\citep{cai2016null,li2016model,wei2017improved,rana2017constraining}
(see \citet{qi2018parameterized} for the latest measurements of
$H(z)$ data). It was found that the cosmic curvature estimated this
way is well consistent with the flat case of $\Omega_k=0$. One
should mention that also other methods to measure the cosmic
curvature have been discussed. \citet{takada2015geometrical}
proposed that combined radial and angular diameter distances from
the BAO can be used to constrain the curvature parameter. The
achievable accuracy of such $\Omega_k$ measurement could be $\Delta
\Omega_k \simeq 10^{-3}$.  \citet{denissenya2018cosmic} used the
measurements of strong lensing time delays and supernova distance to
measure the curvature. In other papers, measurements of $\Omega_k$
are discussed using the mean image separation statistics of
gravitationally lensed quasars \citep{rana2017constraining} or in
the non-flat $\Lambda$CDM inflation model \citep{ooba2018planck}.

The approach of \citet{rasanen2015new} involves strong gravitational
lensing (SGL) systems, where three objects in the Universe (the
source, the lens and the observer) are almost perfectly aligned.
This setting allows testing the type of FLRW metric (and thus
$\Omega_k$) using the distance sum rule \citep{liao2017test}. More
recently, this method has been applied by \citet{xia2017revisiting}
on a SGL sample including 118 galactic-scale systems compiled by
\citet{cao2015cosmology}. In all these previous works, distances
were inferred from type Ia supernova (SN Ia) acting as standard
candles. However, as discussed in the literature
\citep{wei2017improved,li2016model}, using SN Ia data generates
considerable uncertainty of derived $\Omega_k$, due to several
nuisance parameters characterizing the SN light-curves. Moreover,
the limited redshift range of SN Ia, i.e. $z\leq 1.40$ implies that
only a limited number of known SGL systems can be used, and higher
redshift SGL systems cannot be matched with SN Ia
\citep{xia2017revisiting}. Finally, the use of SN Ia data in this
context relies on the so-called Etherington duality principle
connecting luminosity and angular diameter distances. This distance
duality is valid in any metric theory. However, it relies on the
assumption that photons are conserved along the path from the source
to observer, which potentially might be violated
\citep{holanda2011cosmic,cao2011a,cao2011testing,cao2014cosmic}.

It is clear that, for the purpose of implementing the method of
\citet{rasanen2015new}, it would be beneficial to use distance
probes covering higher redshifts thus taking advantage of larger
sample of SGL systems. In this paper we use the latest catalog of
strong gravitational lensing systems \citep{cao2015cosmology} to
extract the distance ratio $\frac{d_{ls}}{d_s}$. Moreover, for the
purpose of distance estimation we use the recently compiled sample
of milliarcsecond compact radio-sources data set comprising 120
intermediate-luminosity quasars calibrated as standard rulers and
covering the redshift range $0.46<z<2.76$ \citep{cao2017ultra}. This
paper is organized as follows: In section \ref{method} we describe
the methodology and the data used in our work. We discuss our
results in section \ref{results1}. Section~\ref{results2}
demonstrates the performance of our method on simulated yields of
future surveys. Finally, the conclusions are presented in section
\ref{conclusion}.

\section{Methodology and data} \label{method}

We assume that in the homogeneous and isotropic universe, its
geometry can be described by the FLRW metric
\begin{equation}
ds^2=-dt^2+\frac{a(t)^2}{1-kr^2}dr^2+a(t)^2r^2d\Omega^2,
\end{equation}
where $k$ represents the spatial curvature. $k=+1, -1, 0$
corresponds to closed, open, and flat universe, respectively and is
related to the curvature parameter as $\Omega_k=-k/H_0^2$.

For a given strong lensing system with the source at redshift $z_s$
and lensing galaxy at redshift $z_l$ the separation of multiple
images depends on the ratio $D_{A}(z_l,z_s)/D_{A}(z_s)$ of
angular-diameter distances between the lens and the source and
between the observer and the source, provided one has a reliable
model for the mass distribution of the lens
\citep{grillo2008cosmological,Biesiada2010,cao2012constraints,cao2015cosmology}.
The dimensionless distance $d(z)$ between the lensing galaxy and the
source is given by
\begin{eqnarray}
d(z_l,z_s)&=&(1+z_s)H_0 D_A(z_l,z_s) \nonumber \\
 &=& \frac{1}{\sqrt{|\Omega_k|}}\left\{
\begin{array}{lll}
\sinh \sqrt{|\Omega_k|}\int^{z_s}_{z_l}\frac{H_0dz'}{H(z')}, &&k<0, \\
\sqrt{|\Omega_k|}\int^{z_s}_{z_l}\frac{H_0dz'}{H(z')}, &&k=0, \\
\sin \sqrt{|\Omega_k|}\int^{z_s}_{z_l}\frac{H_0dz'}{H(z')}, &&k>0,
\end{array}
\right.
\end{eqnarray}
where $\Omega_k=-k/H_0^2$. For convenience, we denote $d_{ls}\equiv
d(z_l,z_s)$, $d_l\equiv d(0,z_l)$ and $d_s\equiv d(0,z_s)$.
According to the distance sum rule \citep{rasanen2015new}, these
three dimensionless distances satisfy the following relation
\begin{equation} \label{smr}
\frac{d_{ls}}{d_s}=\sqrt{1+\Omega_kd_l^2}-\frac{d_l}{d_s}\sqrt{1+\Omega_k d_s^2}.
\end{equation}
Therefore, if the distances $d_l$, $d_s$ and $d_{ls}$ are obtained
from observations, the value of $\Omega_k$ could be directly derived
without the involvement of any specific cosmological model. In this
paper, the distance ratio $d_{ls}/d_s$ is extracted form SGL
systems, while the other two distances, $d_l$ and $d_s$, are
obtained through the angular size measurements of compact structure
in radio quasars calibrated as standard rulers.

\subsection{Strong gravitational lensing systems}

Since the discovery of the first gravitational lens Q0957+561
\citep{walsh19790957+}, and with the increasing size of SGL systems
detected strong lensing has become a serious and important technique
in extragalactic astronomy (exploration of galactic structure)
\citep{ofek2003redshift,cao2016limits} and in cosmology
\citep{Biesiada2010,biesiada2011dark,cao2011testing,cao2011constraints,
cao2012constraints,cao2015cosmology,Lixiaolei2016}. Since early-type
galaxies are more massive and dominate in all strong lensing
surveys, we will consider only SGL systems in which elliptical
galaxies act as lenses. Although the properties of early-type
galaxies, their formation and evolution are still not fully
understood in details, singular isothermal sphere (SIS) and singular
isothermal ellipsoid (SIE) models are commonly used to describe the
mass distribution of lensing galaxies acting as lenses
\citep{koopmans2006the}.

We will consider three types of lens models
which have been extensively used in strong lensing studies.

i) SIS model: Assuming that the elliptical galaxy modeled as singular isothermal sphere (SIS) acts as a lens, the
distance ratio can be expressed as
\begin{equation}\label{SIE_E}
\frac{d_{ls}}{d_s}=\frac{c^2\theta_E}{4\pi \sigma_{SIS}^2}
\end{equation}
The velocity dispersion $\sigma_{SIS}$ reflects the total mass of
the lens (including dark matter). In order to quantify its relation
to the observed velocity dispersion of stars $\sigma_0$, we
introduce a free parameter $f_E=\sigma_{SIS}/\sigma_0$ with the
corresponding prior of $0.8<f_E^2<1.2$
\citep{ofek2003redshift,cao2012constraints}.

ii) Power-law spherical model: We assume that the total
mass density profile follows spherically symmetric power-law
distribution $\rho\sim r^{-\gamma}$, where $r$ is the spherical
radius from the center of the lensing galaxy. Combined with the
spherical Jeans equation, one can express observational value of the
angular-diameter distance ratio as
\citep{koopmans2006the,cao2015cosmology}
\begin{equation} \label{sigma_gamma}
\frac{d_{ls}}{d_s}=\frac{c^2\theta_E}{4\pi
\sigma_{ap}^2}\left(\frac{\theta_{ap}}{\theta_E}\right)^{2-\gamma}f^{-1}(\gamma),
\end{equation}
where
\begin{equation}
f(\gamma)=-\frac{1}{\sqrt{\pi}}\frac{(5-2\gamma)(1-\gamma)}{3-\gamma}\frac{\Gamma(\gamma-1)}{\Gamma(\gamma-3/2)}\left[\frac{\Gamma(\gamma/2-1/2)}{\Gamma(\gamma/2)}\right]^2.
\end{equation}
Note that $\sigma_{ap}$ is the velocity dispersion of the lens obtained inside
an aperture radius $\theta_{ap}$ (more precisely,
luminosity-averaged line-of-sight velocity dispersion).
Moreover, we allow the power-law index to vary with redshift $\gamma(z) = \gamma_0 + \gamma_1 z$ in order
to account for the possible evolution of the mass density profile.
The analysis performed by \cite{ruff2011sl2s} using the same parametrization of $\gamma(z)$
revealed the following results: $\gamma(z_l) =
2.12^{+0.03}_{-0.04} - 0.25^{+0.10}_{-0.12} \times z_l+
0.17^{+0.02}_{-0.02}$, where the last term denotes the Gaussian scatter about that linear relation. Similar works aimed at
establishing the evolution of mass density profile with different
samples of lens galaxies have also been attempted in the literature
\citep{koopmans2006the,bolton2012the}. In our
analysis, $\gamma_0$ and $\gamma_1$ are treated as free
parameters together with cosmological ones.

iii) Extended power-law model: As a third lens model we consider a more
complex one, which allows the luminosity density profile
$\nu(r)$ differ from the total-mass density profile $\rho(r)$, both
of them modeled by corresponding power-laws $\rho \sim r^{-\alpha}$,
$\nu \sim r^{-\delta}$. Moreover, we consider the anisotropic
distribution of three-dimensional velocity dispersion, i.e. we admit
that radial and tangential velocity dispersions might be different.
This is captured in the anisotropy parameter: $\beta(r) = 1 -
{\sigma^2_t} / {\sigma^2_r}$. In this model the observational value
of the distance ratio is given by \citep{schwab2009galaxy}
\begin{eqnarray}\label{sigma_alpha_delta}
\nonumber
\frac{d_{\rm ls}}{d_{\rm s}}&=& \left(\frac{c^2}{4\sigma_{0}^2}\theta_{\rm E}\right)\frac{2}{\sqrt{\pi}(\xi-2\beta)} \left( \frac{\theta_{\rm ap}}{\theta_{\rm E}}\right)^{2-\alpha}\\
&\times&\left[\frac{\lambda(\xi)-\beta\lambda(\xi+2)}{\lambda(\alpha)\lambda(\delta)}\right]
\frac{\Gamma{\left(\frac{3-\xi}{2}\right)}}{\Gamma{\left(\frac{3-\delta}{2}\right)}}~,
\end{eqnarray}
where $\xi=\alpha+\delta-2$, and
$\lambda(x)=\Gamma(\frac{x-1}{2})/\Gamma(\frac{x}{2})$. Following
the analysis of \citet{cao2015cosmology}, we will model the
anisotropy parameter by a Gaussian distribution $\beta=0.18\pm0.13$,
based on the well-studied sample of nearby elliptical galaxies.

The methodology described above is implemented to the sample of 118
galactic scale SGL systems from the Sloan Lens ACS Survey (SLACS),
BOSS emission-line lens survey (BELLS), Lens Structure and Dynamics
(LSD) and Strong Lensing Legacy Survey (SL2S) assembled by
\cite{cao2015cosmology}. Moreover, recent analysis of
different sub-samples lenses defined according to their redshifts and
velocity dispersions (i.e. effectively according to their masses) showed that it is necessary to investigate separately
low-, intermediate- and high-mass galaxies \citep{cao2016limits}.
Since the redshift range covered by lenses was
not large enough to display any noticeable differences, the division
according to velocity dispersion turned out to be more
discriminative. Therefore, we also divide the SGL sample into three
sub-samples based on lens velocity dispersion: $\sigma_{ap}\leq 200
\; \rm{km~s^{-1}}$ (low-mass galaxies), $200 \;
\rm{km~s^{-1}}<\sigma_{ap}\leq 300 \; \rm{km~s^{-1}}$
(intermediate-mass galaxies) and $\sigma_{ap}>300 \; \rm{km~s^{-1}}$
(high-mass galaxies), and study the effect on the measurement of
curvature caused by such division.

\subsection{Radio quasar data}

Currently, the possibility of using compact radio sources to study
cosmological parameters and physical properties of AGNs became very
attractive in the literature
\citep{jackson1997deceleration,vishwakarma2001consequences,
lima2002dark,zhu2002cardassian,chen2003cosmological}. More
interestingly, the advantage of radio quasars, compared with other
standard rulers including baryon acoustic oscillations (BAO)
\citep{percival2010baryon}, galaxy clusters
\citep{bonamente2006determination}, and strong lensing systems
\citep{cao2012constraints,cao2015cosmology}, lies in the fact that
is that quasars are observed at much higher redshifts. Our procedure
follows the phenomenological model originally proposed in
\cite{gurvits1994apparent} and later investigated in
\cite{gurvits1998angular}, which quantifies the luminosity and
redshift dependence of the linear sizes of quasars as
\begin{equation}
l_m=lL^{\beta}(1+z)^n,
\end{equation}
where $\beta$ and $n$ are two constant parameters quantifying the
``angular size-luminosity'' and ``angular size-redshift'' relations.
Therefore, constraints on $\beta$ and $n$ will help us to
differentiate between sub-samples. In particular the sub-sample of
compact radio sources fulfilling $\beta=n=0$ could act as standard
rulers. Considering the VLBI visibility data for 337 active galactic
nuclei (AGN), \cite{gurvits1994apparent} provided a rough estimation
of the dependence of apparent angular sizes on the luminosity and
redshifts: $\beta\sim0.26$, and $n\sim-0.30$, under assumption of
homogeneous and isotropic universe without cosmological constant.
Regression analysis of the restricted sample with spectral index
($-0.38\leq \alpha\leq 0.18$) and total luminosity ($Lh^2\geq
10^{26}$ W/Hz), gave a value of $\beta\sim0.37$ and $n\sim-0.58$
\citep{gurvits1998angular}. Following this direction, in the
framework of concordance $\Lambda$CDM cosmology
\cite{Caoexploring2015} in attempt to determine intrinsic linear
sizes of compact structure in 112 radio quasars found a substantial
evolution of linear sizes with luminosity ($\beta\sim0.17$) and much
smaller cosmological evolution of the linear size ($|n|\simeq
10^{-2}$). More recently, the use of intermediate-luminosity quasars
($10^{27}$ W/Hz $<L<10^{28}$ W/Hz) as potential cosmological tracers
was studied in detail in \cite{cao2017measuring}. The sample of 613
milliarcsecond ultra-compact radio sources based on a 2.29 GHz VLBI
all-sky survey \citep{Kellermann93,gurvits1994apparent}, was divided
into different sub-samples, according to their optical counterparts
and luminosity: low, intermediate, and high-luminosity quasars.
Luminosity selection as well as $D_A(z)$ assessments necessary for
building the sample were performed without pre-assuming a
cosmological model but basing on the $D_A(z)$ reconstruction from
$H(z)$ data obtained from cosmic chronometers
\citep{2002Jimenez,Moresco:2012jh}. Further detailed analysis
revealed that only intermediate-luminosity quasars (ILQSO) show
negligible dependence on both redshifts $z$ and intrinsic luminosity
$L$ ($|n|\simeq 10^{-3}$, $\beta\simeq 10^{-4}$), which makes them a
fixed comoving-length standard ruler. Subsequently,
\cite{cao2017ultra} used an improved cosmological-model-independent
method to calibrate the linear sizes of ILQSO as $l_m=11.03\pm0.25$
pc at 2.29 GHz. Cosmological application of this data set
\citep{cao2017ultra} resulted with stringent constraints on both the
matter density parameter $\Omega_m$ and the Hubble constant $H_0$,
in a very good agreement with recent \textit{Planck} 2015 results
\citep{ade2016planck}. The exploration of other cosmological models
in light of this observational data has been broadly studied in the
literature \citep{Li2017,Qi2017,Ma2017,Zheng2017,Xu2018,Cao2018}. In
the context of our study, it is reasonable to ask whether
uncertainties in the standard ruler assumption generate additional
systematic errors on the final result. In order to address this
issue, we performed a sensitivity analysis by applying Monte Carlo
simulations in which $\beta$, $n$ parameters are characterized by
Gaussian distributions: $\beta=0.00\pm0.05$ and $n=0.00\pm0.05$,
respectively, while the uncertainty of the linear size scaling
factor was taken into account with a Gaussian distribution as
$l=11.03\pm0.25$ pc. In result the effects of uncertainties of $l$,
$\beta$ and $n$ turned out to be negligible, especially for the
reconstruction of $d(z)$ function from the quasar data.

We will use the above mentioned calibrated value of the linear size
$l_m$ to calculate the angular diameter distances to ILQSO sample
\citep{cao2017ultra}
\begin{equation}
D_A(z)=\frac{l_m}{\theta(z)}.
\end{equation}
Then one can use angular diameter distances to the quasars to
obtain the dimensionless distances to the lenses $d_l$
and to the sources $d_s$ (for each SG system), as
\begin{equation}
d_l=\frac{H_0}{c}(1+z_l)D_A(z_l); \ d_s=\frac{H_0}{c}(1+z_s)D_A(z_s).
\end{equation}
For this purpose we assume the value of the Hubble constant $H_0=67.8 \;
\rm{kms}^{-1}\rm{Mpc}^{-1}$ after \textit{Planck} Collaboration XIII
(2015).

\begin{figure}
\centering
\includegraphics[width=1.0\hsize]{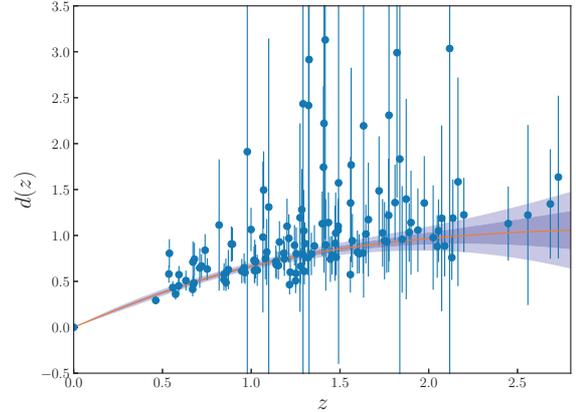}
\caption{Reconstruction of $d(z)$ from the quasar sample. Shaded areas represent $68\%$ and $95\%$ confidence level of the reconstruction.}\label{GP}
\end{figure}

The problem is that in order to use ILQSO distances in
Eq.~(\ref{smr}) one should identify intermediate luminosity quasars
at redshifts equal to (or at least very close to) the redshifts of
sources and lenses in SGL systems from which distance ratio is
inferred. This is not possible, but fortunately, a model-independent
method of Gaussian processes (GP) \citep{seikel2012reconstruction},
can be employed to reconstruct the dimensionless comoving distance
from the data straightforwardly, without any parametric assumption
regarding cosmological model. In this process, the values of
the reconstructed function evaluated at any two different points $z$
and $\tilde{z}$, are connected by a covariance function
$k(z,\tilde{z})$. In this paper, we take the squared exponential
form for covariance function $k(z,\tilde{z})$, which depends only on
two hyper parameters $\sigma_f$ and $\ell$
\begin{equation}
k(z,\tilde{z})=\sigma_f^2\exp\left(-\frac{(z-\tilde{z})^2}{2\ell^2}\right).
\end{equation}
The values of these hyper parameters can be optimized by the GP
itself via the observed data. The GP approach has been used in
various studies
\citep{seikel2012using,yang2015reconstructing,li2016constructing,qi2016testing,zhang2016test}.
Therefore, we use GP to reconstruct the profile of $d(z)$ from the
quasar sample as shown in Fig. \ref{GP}. Obviously, the anchoring
point $d(0)=0$ makes the uncertainty of the reconstructed $d(z)$
smaller at low redshifts. This reconstruction can be done up to the
redshift $z=2.8$. Considering the redshift range of lenses
$0.075\leq z_l \leq 1.004$ and sources $0.196\leq z_s \leq 3.595$ in
the original sample of 118 strong lenses, SGL systems with $z_s>2.8$
should be excluded. This selection leaves us with 106 SGL systems
and we performed our study on this sample, in which low-mass lensing
galaxies contribute 20 data points ($0.075\leq z_l \leq 0.682$),
intermediate-mass galaxies contribute 73 data points ($0.082\leq z_l
\leq 0.938$), and high-mass galaxies contribute 13 data points
($0.164\leq z_l \leq 1.004$).

\begin{figure*}
\centering
\includegraphics[width=1.0\hsize]{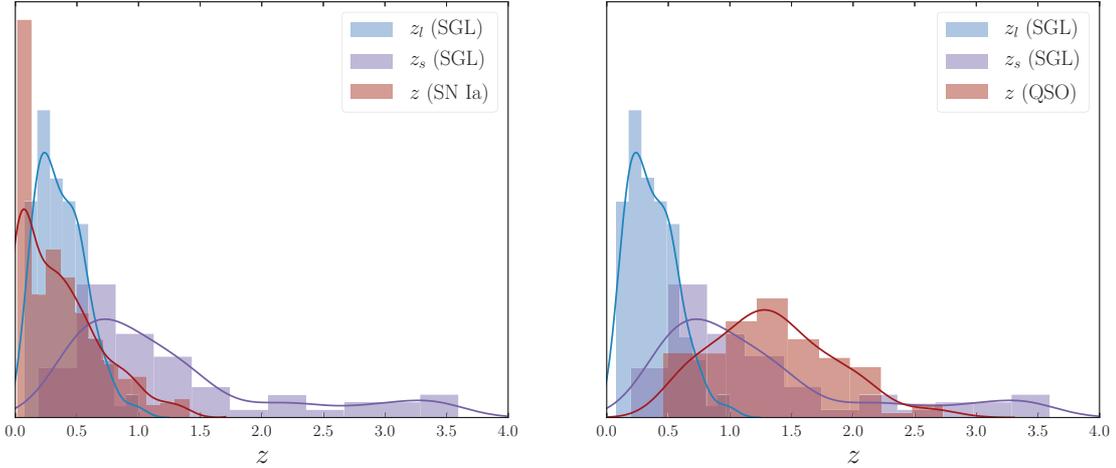}
\caption{ Redshift distribution of SGL systems and different cosmological
probes used to assess distances. One can see that QSO sample has more fair coverage of
the redshift range of SGL sources than SN Ia.}\label{redshift_range}
\end{figure*}

Fig.~\ref{redshift_range} shows the redshift coverage of strong
lensing systems, SN Ia and ILQSO. Compared with the previous works
using the luminosity distances $D_L(z)$ of SN Ia
\citep{xia2017revisiting}, one can see that the intermediate
luminosity quasars have better coverage of redshifts corresponding
to the sources in SGL systems.

\section{Results and discussions} \label{results1}

The $\Omega_k$ parameter was determined by maximizing the likelihood
${\cal L} \sim \exp{(-\chi^2 / 2)}$ where the chi-square function
was defined as:
\begin{equation}
\chi^2(\textbf{p},\Omega_k)=\sum_{i=1}^{106}\frac{\left(D_{th}(z_i;\Omega_k)-D_{obs}(z_i;\textbf{p})\right)^2}{\sigma_D(z_i)^2},
\end{equation}
where $D=d_{ls}/d_s$ and its theoretical value (dependent on
$\Omega_k$) is given by the distance sum rule Eq.~(\ref{smr}) and
its observational counterpart is Eq.~(\ref{SIE_E}), Eq.~(\ref{sigma_gamma}) or
Eq.~(\ref{sigma_alpha_delta}) depending on the lens model adopted.
Parameters of the lens model are denoted as $\textbf{p}$. We fit both $\Omega_k$ and nuisance parameters
$\textbf{p}$ and later we marginalize over the latter. Maximization
was performed using the emcee Python module \citep{foreman2013emcee}
based a Markov chain Monte Carlo (MCMC) code. Two factors contribute
to the uncertainty of $D$: uncertainty from SGL systems and
uncertainty from the distances reconstructed from ILQSO. We assume
that being uncorrelated they add in quadrature:
$\sigma_D^2=\sigma_{SGL}^{2}+\sigma_{QSO}^2$. Term from SGL systems
comes from the uncertainty of velocity dispersion measurements
(reported in the data set) and from the Einstein radius
determination which we assume at the level of $5 \%$. Exact formulae
for these uncertainties in the lens models considered can be found
in \cite{cao2012constraints,cao2015cosmology,cao2016limits} respectively. Part of the
uncertainty coming from quasars $\sigma_{QSO}$ is determined as a
width of the uncertainty strip of GP-reconstructed distance for a given redshift $z_{(l,s),i}$.

\subsection{SIS model}

\begin{figure*}
\centering
\includegraphics[width=8cm,height=8cm]{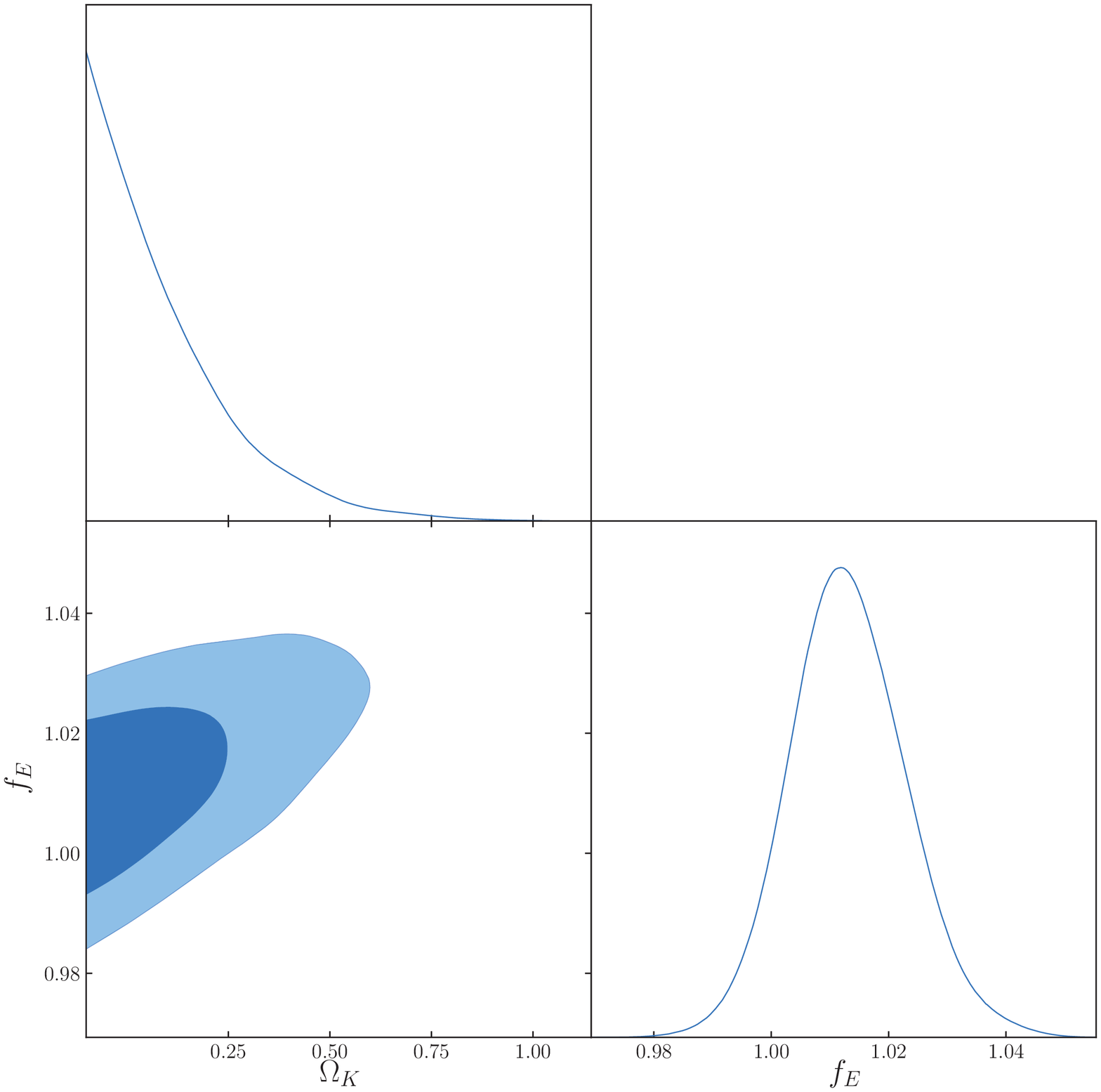}
\includegraphics[width=8cm,height=8cm]{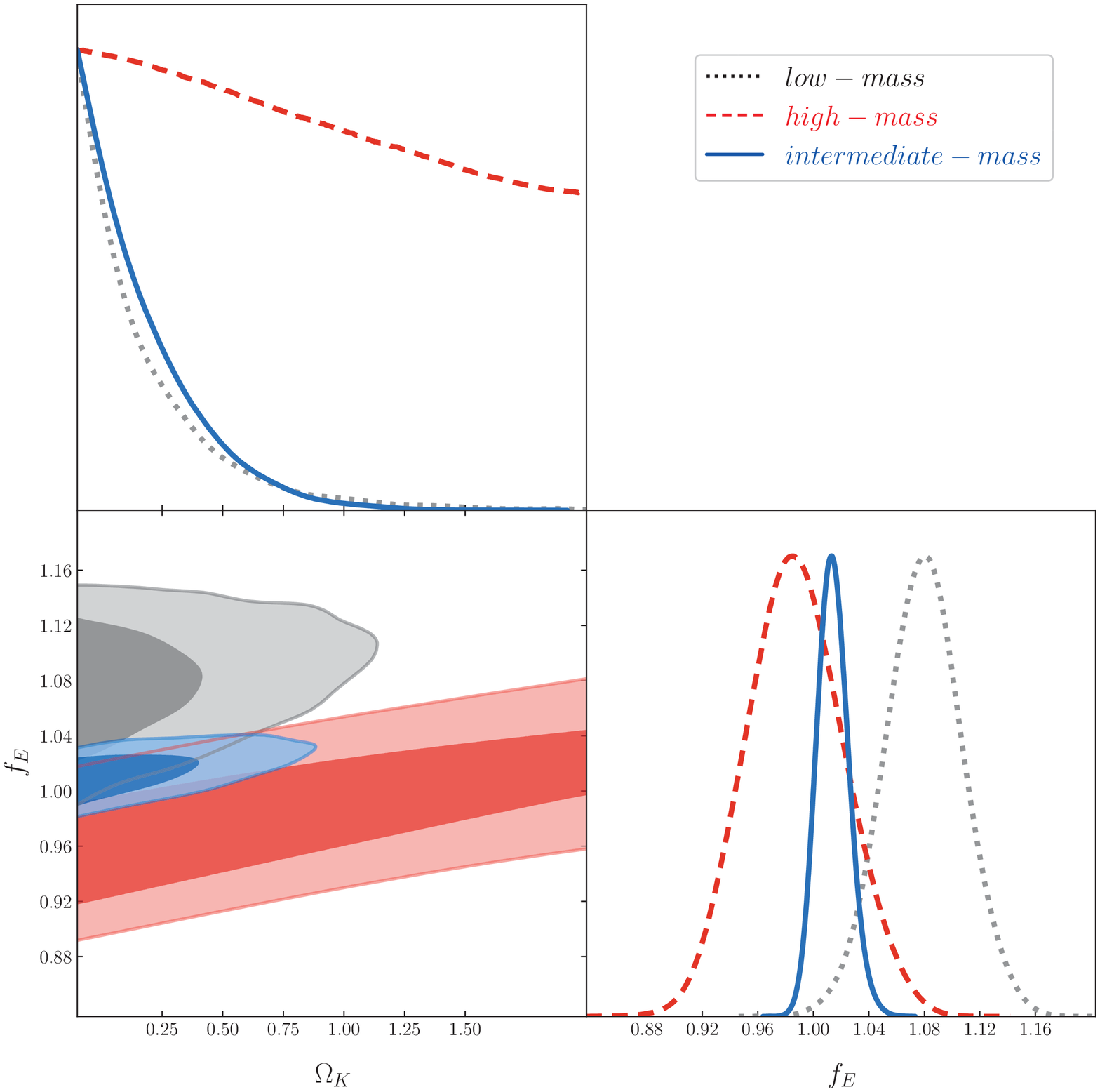}
\caption{Left: 2-D distributions of cosmic curvature $\Omega_k$ and
SGL parameters ($f_E$) constrained from the 106 SGL sample in the
case of SIS lens model. $1\sigma$ and $2\sigma$ uncertainty contours
are shown together with 1-D distributions of respective parameters
marginalized over the remaining ones. Right: The results obtained
from three sub-samples of SGL systems selected according to their
velocity dispersions.}\label{SIE_fig}
\end{figure*}

\begin{table*}
\begin{center}
\begin{tabular}{|c| c| c|c|c|}
\hline
         & low-mass & intermediate-mass & high-mass & full sample \\
\hline
$\Omega_k$& $<0.231$ & $<0.227$ & ---& $<0.136$ \\
\hline
$f_E$ & $1.08\pm0.028$& $1.014\pm 0.011$& $0.987\pm0.035$& $1.0132^{+0.0092}_{-0.010}$\\
\hline
\end{tabular}
\caption{ Results from the SIS model: the best-fitted values of
$\Omega_k$ and $f_E$ with $68\%$ confidence level for different
samples.} \label{SIE_table}
\end{center}
\end{table*}

Working on the SIS lens model, we obtained the results displayed in
Fig.~\ref{SIE_fig} and Table~\ref{SIE_table}. The most stringent
constraint on the cosmic curvature $\Omega_k$ was obtained from the
full sample of SGL systems. Low and intermediate-mass lenses gave
less stringent constraints. However, the flat Universe is supported
in all these cases. Due to the limited sample size, high-mass sample
was too small to give reasonable constraint. Concerning the $f_E$
parameter, its best-fitted value is different across the
sub-samples. However, high and intermediate-mass lens samples agree
with each other and with the full sample in the sense that their
$1\sigma$ regions overlap. Low-mass sample differs from the rest up
to $2\sigma$ interval. More interestingly, we find the SIS model
($f_E=1$) is well consistent with results obtained in low-mass and
intermediate-mass (low and intermediate velocity dispersion)
galaxies. Consequently, our results imply that consideration of
sub-samples defined according to velocity dispersions is necessary
in SGL analysis.

\subsection{Power-law spherical model}

\begin{figure*}
\centering
\includegraphics[width=8cm,height=8cm]{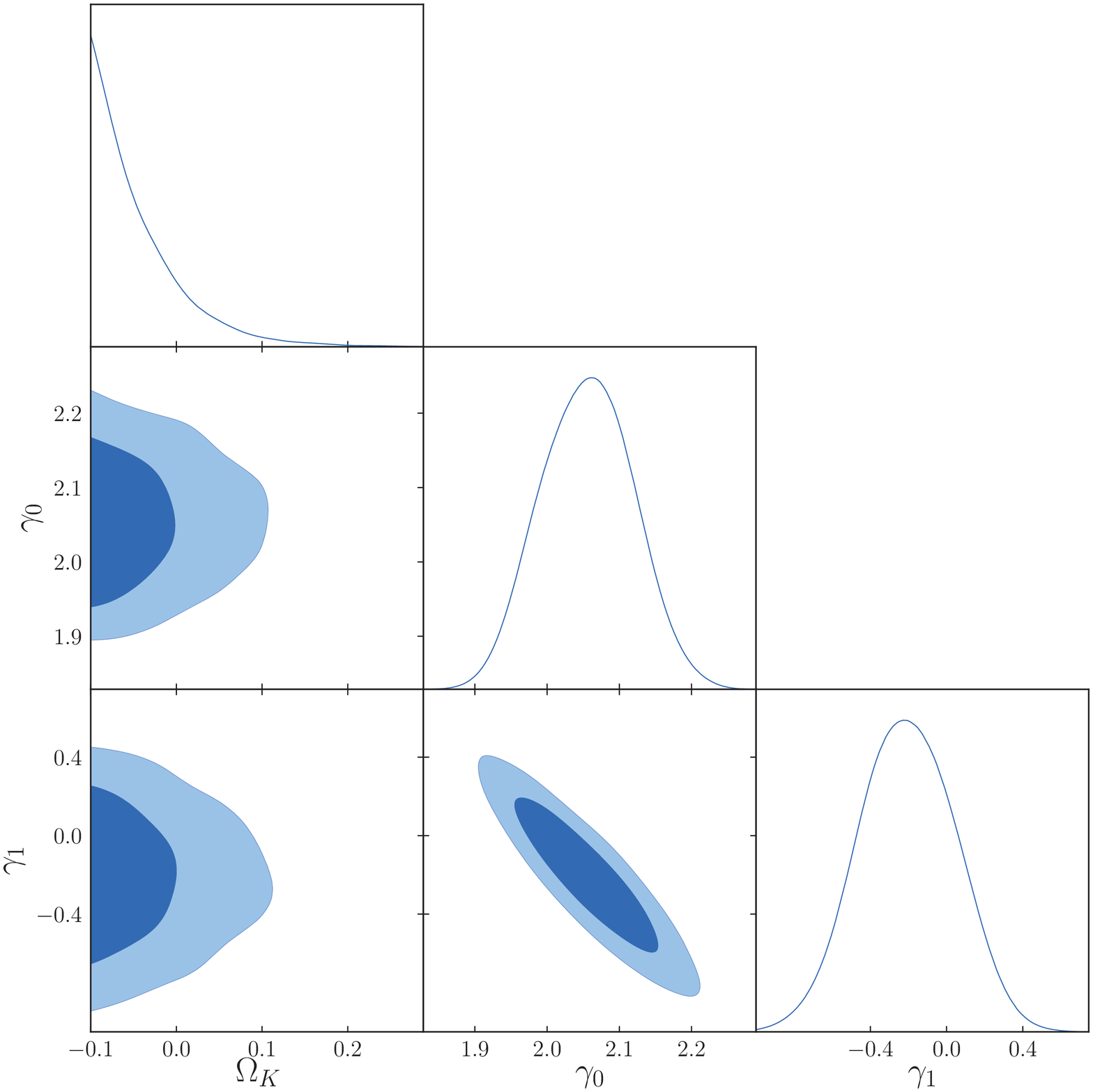}
\includegraphics[width=8cm,height=8cm]{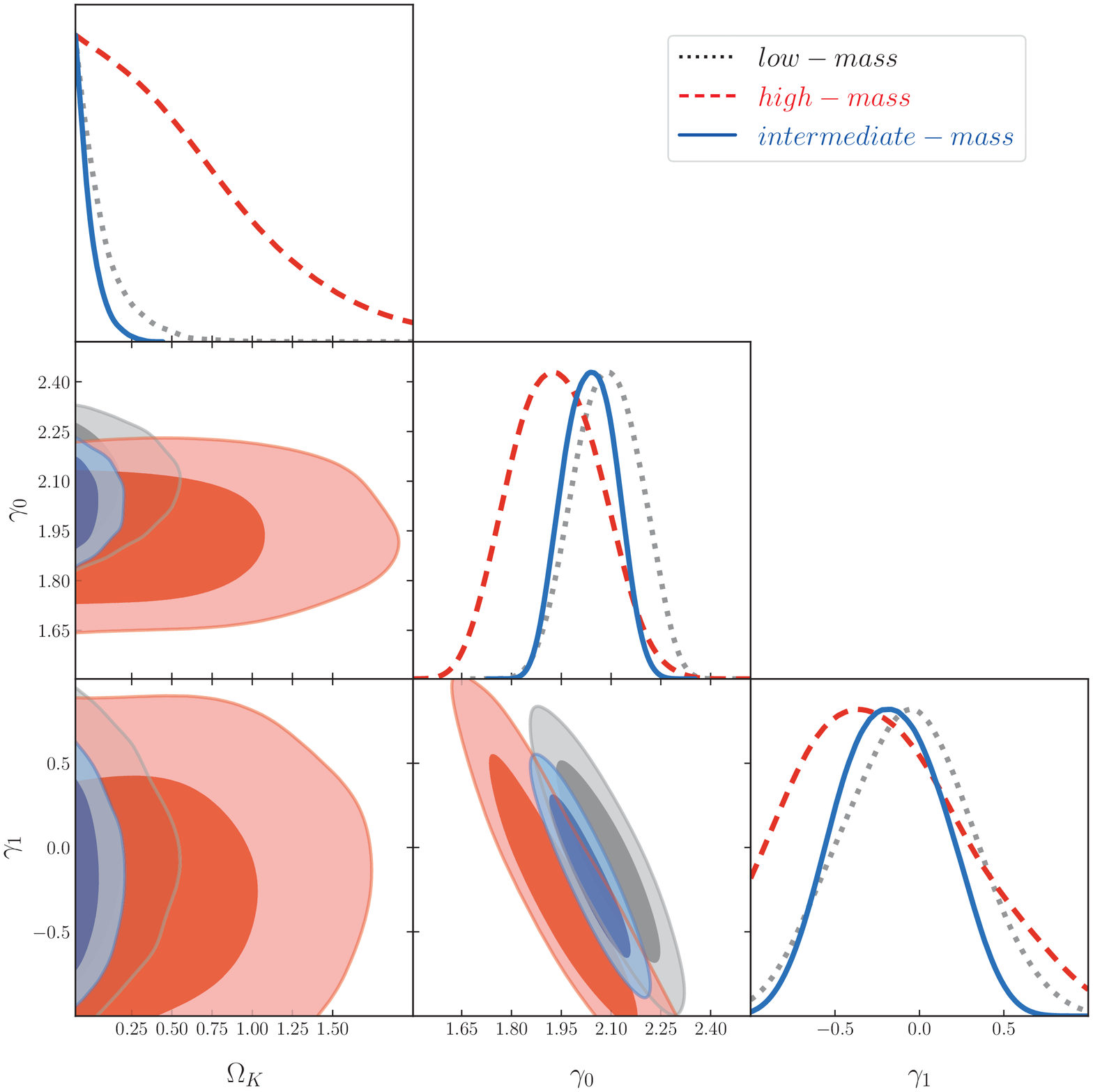}
\caption{Left: 2-D distributions of cosmic curvature $\Omega_k$ and
SGL parameters ($\gamma_0$, $\gamma_1$) constrained from the 106 SGL
sample in the case of spherically symmetric power-law lens model.
$1\sigma$ and $2\sigma$ uncertainty contours are shown together with
1-D distributions of respective parameters marginalized over the
remaining ones. Right: The results obtained from three sub-samples
of SGL systems selected according to their velocity
dispersions.}\label{SIS_fig}
\end{figure*}

\begin{table*}
\begin{center}
\begin{tabular}{|c| c| c|c|c|}
\hline
         & low-mass & intermediate-mass & high-mass & full sample \\
\hline
$\Omega_k$& $<0.0883$ & $<-0.0012$ & $<0.717$& $<-0.0327$ \\
\hline
$\gamma_0$ & $2.089\pm0.098$& $2.036\pm0.077$& $1.94\pm0.13$& $2.054\pm0.064$ \\
\hline
$\gamma_1$ & $-0.07\pm0.37$& $-0.17\pm0.31$ & $-0.19^{+0.35}_{-0.61}$& $-0.20\pm0.26$\\
\hline
\end{tabular}
\caption{ Results from the power-law spherical model: the
best-fitted values of $\Omega_k$, $\gamma_0$ and $\gamma_1$ with
$68\%$ confidence level for different samples.} \label{SIS_table}
\end{center}
\end{table*}

In the case of spherically symmetric power-law lens model (allowing
for the evolution of the power-law exponent) we obtained the results
shown in Fig.~\ref{SIS_fig} and Table.~\ref{SIS_table}. For the full
sample, the central fits are $\Omega_k<-0.0327$,
$\gamma_0=2.054\pm{0.064}$ and $\gamma_1=-0.20\pm 0.26$.
Marginalized distribution of $\Omega_k$ implies that, flat Universe
is still compatible with the data within $2\sigma$ confidence
interval. However, a closed universe seems to be more favored, which
is also in good agreement with the results of previous works
\citep{rasanen2015new,xia2017revisiting}. It is, however, different
from clearly preferred by the CMBR experiments
\citep{ade2016planck}. Even though our constraint on the cosmic
curvature is not improved over the previous works
\citep{rasanen2015new,xia2017revisiting}, yet our approach more
fairly covers the redshift range of sources whose angular diameter
distances are being used. Let us stress again, that we have not only
constrained the cosmic curvature, but also the evolution of slope
factor in the mass density profile of lensing galaxies. Our results
suggest that the total density profile of early-type galaxies have
become slightly shallower over cosmic time, but it is still
consistent with the standard SIS model ($\gamma_0=2$, $\gamma_1=0$)
within 1$\sigma$ uncertainty.

The analysis performed on sub-samples defined by the lens velocity
dispersion reveals that, even though the best fitted values of slope
parameters ($\gamma_0$, $\gamma_1$) are different, they are all
consistent with each other and with the full sample within $1\sigma$
regions. We note that the ranges of $\gamma$ parameters, which
quantify the corresponding quantities of relatively high-mass
galaxies, are very close to the estimates obtained for low-mass and
intermediate-mass elliptical galaxies. Therefore, our analysis
indicates the mass distribution in massive lensing galaxies can be
effectively described by the power-law spherical model. Moreover,
one can clearly see from Table 2 that in the framework of power-law
spherical lens model, all of the three sub-samples of galaxies will
provide fitting results in one direction. The most stringent
constraint on $\Omega_k$ is provided by the full sample, i.e., a
closed universe is strongly supported by the current observational
SGL data.

\begin{figure*}
\centering
\includegraphics[width=8cm,height=8cm]{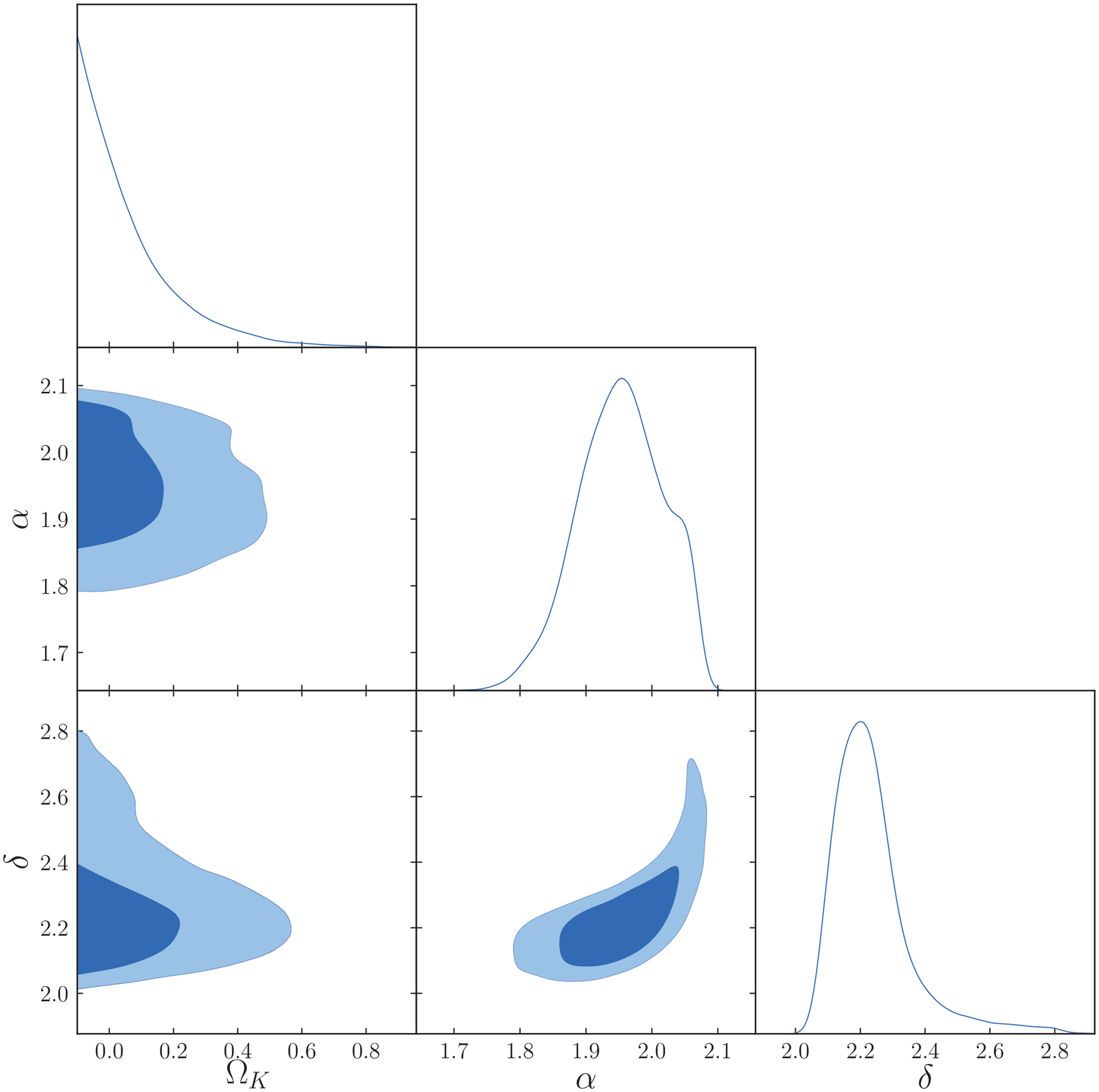}
\includegraphics[width=8cm,height=8cm]{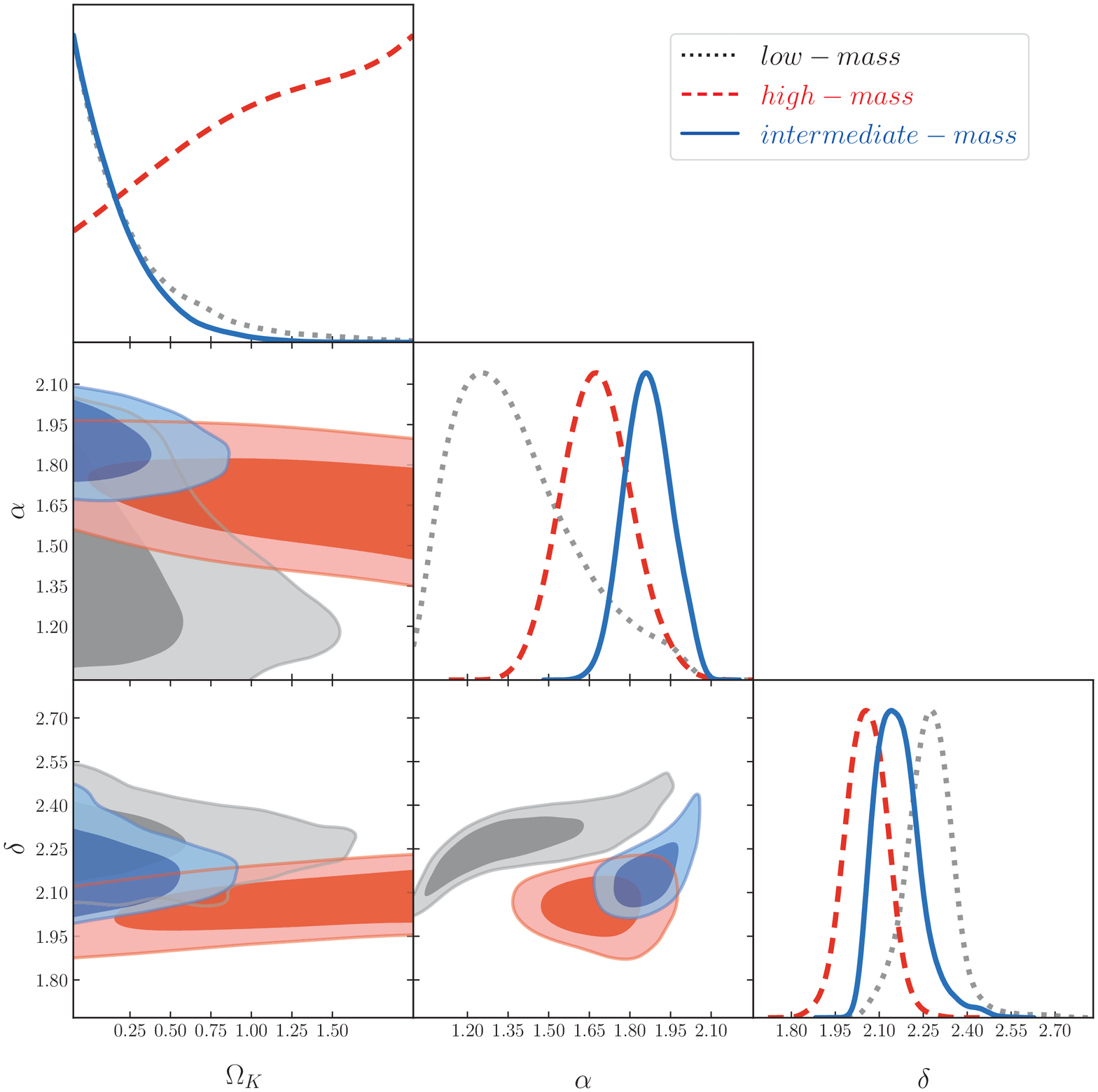}
\caption{2-D distributions of cosmic curvature $\Omega_k$ and SGL
parameters ($\alpha$, $\delta$) constrained from the 106 SGL sample
in the case of generalized lens model where stellar and total mass
components follow different power-laws. $1\sigma$ and $2\sigma$
uncertainty contours are shown together with 1-D distributions of
respective parameters marginalized over the remaining ones. Right:
The results obtained from three sub-samples of SGL systems selected
according to their velocity dispersions. }\label{ESIE_fig}
\end{figure*}

\begin{table*}
\begin{center}
\begin{tabular}{|c| c| c|c|c|}
\hline
         & low-mass & intermediate-mass & high-mass & full sample \\
\hline
$\Omega_k$& $<0.324$ & $<0.22$ & $>0.790$& $<0.0858 $ \\
\hline
$\alpha$ & $1.38^{+0.12}_{-0.29}$& $1.866\pm0.086$& $1.68\pm0.13$& $1.96^{+0.086}_{-0.06}$ \\
\hline
$\delta$ & $2.274\pm0.086$& $2.164^{+0.055}_{-0.089}$ & $2.06\pm0.071$& $2.27^{+0.035}_{-0.16}$\\
\hline
\end{tabular}
\caption{ Results from the extended power-law model: the best-fitted
values of $\Omega_k$, $\alpha$ and $\delta$ with $68\%$ confidence
level for different samples.} \label{ESIE_table}
\end{center}
\end{table*}

\subsection{Extended power-law lens model}

In Fig.~\ref{ESIE_fig}, we show the 2-D distributions with
corresponding $1\sigma$ and $2\sigma$ contours for the $\Omega_k$
and lens model parameters in the case where luminosity and total
mass follow different power-law distributions.
Table~\ref{ESIE_table} lists the best-fitted values of the
parameters with $68\%$ confidence level for different sub-samples.
From the fitted range of $\alpha$ and $\delta$ parameters, one can
notice that the luminosity density profile of elliptical galaxies is
different from the total mass density profile ($\alpha \neq
\delta$). Therefore, the issue of mass density profile in the
early-type galaxies is still a critical one that needs to be
investigated further \citep{cao2016limits}. In the framework of
extended power-law lens model, one could obtain a constraint on the
curvature $\Omega_k<0.0858$ at 68\% C.L. from the full sample.
Therefore, a universe with zero curvature (spatially flat geometry)
is strongly supported by the available observations. This is the
most unambiguous result of the current data set.

Concerning the sub-samples, we find that the best-fitting values of
$\Omega_k$ obtained with massive galaxies are significantly
different from the corresponding results with low-mass and
intermediate-mass elliptical galaxies (which agree very well with
the spatial flatness of the Universe). More specifically, the
high-mass sub-sample did not give an upper bound on $\Omega_k$ as
usual, but a lower bound at $1\sigma$ confidence level. However, a
zero value of $\Omega_k$ is still included in $2\sigma$ confidence
level. On the other hand, there also exists a tension between the
values of lens model parameters obtained from different sub-samples
(even beyond the $2\sigma$ confidence interval in $\alpha - \delta$
plane), which supports the scheme of treating separately sub-samples
with different velocity dispersion. Such conclusion has been first
noticed and discussed in \citet{cao2016limits}. Moreover, high-mass
lenses provide clearly different constraints on the slope parameters
($\alpha, \delta$) compared to their low-mass counterparts. Such
findings can be interpreted as neither of the dark matter and
stellar components is well approximated by an isothermal profile,
although they seem to work well together in building a nearly
isothermal total density profile \citep{treu2010strong}. More
specifically, this tendency can also be understood as the result of
different dark matter density profiles, which could fall off more
steeply than the stellar mass in the inner region of high-mass
lensing galaxies. Our results are in perfect agreement with the
recent analysis of massive early-type galaxies from large
cosmological hydrodynamical simulations, which indicated that the
dark matter and total density profiles tend to become slightly
steeper at low redshifts, while the stellar one does not vary
significantly \citep{peirani2018total}.

\begin{figure}
\centering
\includegraphics[width=8cm,height=6cm]{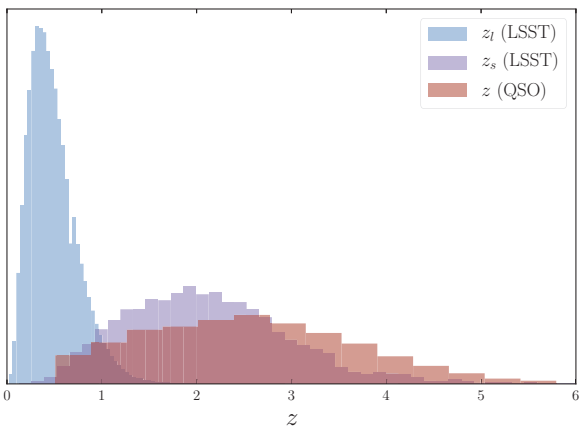}
\caption{ Redshift distribution of the simulated SGL and QSO
sample.} \label{Simulated_redshift}
\end{figure}

\begin{figure}
\centering
\includegraphics[width=8cm,height=6cm]{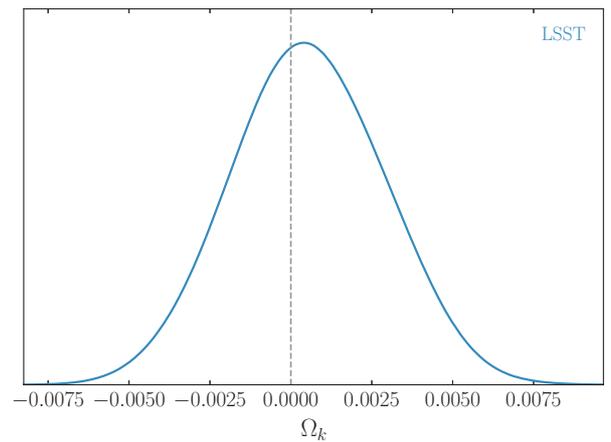}
\caption{Constraints on the cosmic curvature achievable in the
future surveys. Forecast based on simulated strong lensing and
quasar data.} \label{Simulated_result}
\end{figure}

\section{Monte Carlo Simulated Mock Sample of SGL and QSO systems } \label{results2}

The ongoing and future massive surveys like LSST or DES will provide
us the opportunity to discover up to three orders of magnitude more
galactic-scale strong lenses that are currently known. For instance,
according to the prediction of \cite{collett2015the}, the
forthcoming Large Synoptic Survey Telescope (LSST) survey may
potentially discover 120000 lenses for the most optimistic scenario.
Therefore, we can ask the question: how precise can one determine
the cosmic curvature with such a significant increase of the number
of strong lensing systems? On the other hand, the uncertainty of the
angular size of compact structure in radio quasars will be
significantly reduced by both current and future VLBI surveys based
on better uv-coverage \citep{pushkarev2015milky}. Consequently, one
can expect to have a better distance information from the ILQSO
standard ruler approach with future VLBI data.

For the purpose of our simulation we have chosen a fiducial
cosmological model, which was the flat ($\Omega_k=0$) $\Lambda$CDM
with $\Omega_m=0.30$ and $H_0=70 ~\rm{km/s/Mpc}$.
 For the quasar data to be observed by future VLBI
surveys, the mock ``$\theta - z$" data is generated with fixed
linear size of ILQSOs equal to $l_m=11.03$ pc. We have simulated 500
intermediate-luminosity quasars in the redshift range $0.50<z<6.00$,
for which the fractional uncertainty of the angular size ``$\theta$"
was taken at a level of 3\%. The redshift distribution of the
sources observed follows the luminosity function obtained from a
combination of SDSS and 2dF (2SLAQ) surveys, the bright and
faint-end slopes of which agree very well with those in the
bolometric luminosity function. See \citet{cao2017ultra} for more
details of the specific procedure of QSO simulation.

For the strong lensing data, we use the realistic simulation of
120000 events with elliptical galaxies acting as lenses in the
forthcoming LSST survey \citep{collett2015the}. The simulation codes
are available on the github.com/tcollett/LensPop. Considering the
recent result that only the medium-mass elliptical galaxies are
consistent with the singular isothermal sphere case within 1$\sigma$
uncertainty \citep{cao2016limits}, we restrict the velocity
dispersions of lensing galaxies to the intermediate range:
$200~\mathrm{km/s}<\sigma_{ap}<300~\mathrm{km/s}$ and obtain 16000
strong lensing systems meeting the redshift criterion
$0.5<z_l<z_s<6.0$ in compliance with QSO data used in parallel. In
deriving the distances $d_l$ and $d_s$ from QSOs, we use the
selection criterion $|z_{SGL}-z_{QSO}| \leq 0.005$.
Fig.~\ref{Simulated_redshift} shows the redshift coverage of the
simulated QSO and SGL sample. Following the analysis of
\citet{cao2016limits}, the fractional uncertainty of the Einstein
radius and the observed velocity dispersion are respectively taken
at the level of 1\% and 10\%. The posterior probability density for
$\Omega_k$ is shown in Fig.~\ref{Simulated_result}. It should be
emphasized that the central value of this fit reflects the fiducial
model assumed i.e. flat $\Lambda$CDM. It would of course be
different for a different choice of a fiducial model . However, the
point is how precisely can future observational data determine the
cosmic curvature and this is not so much dependent on the
cosmological model used in simulations of mock catalog. We find that
with about 16000 strong lensing events combined with the distance
information provided by 500 compact radio quasars, one can constrain
the cosmic curvature with an accuracy of $\Delta \Omega_k\simeq
10^{-3}$, which is comparable to the precision of \textit{Planck}
2015 results.

\section{Conclusions} \label{conclusion}

The distance sum rule method applied to strong lensing systems is
becoming an important astrophysical tool for probing the curvature
of the universe without assuming any fiducial cosmological model. In
this paper, attempting to alleviate the shortcomings of using SN Ia
as distance indicators in the sum rule, we turned to the VLBI
observations of milliarcsecond compact structure in
intermediate-luminosity quasars covering the redshift range $0.46 <
z < 2.76$, which have recently been demonstrated to be
standardizable rulers. Providing a better redshift coverage of SGL
systems they were used for distance inferences without pre-assuming
any particular parameterized cosmological model. Moreover, in order
to better assess the effect of lens model on measuring the cosmic
curvature, we used three lens models: i) SIS model ii) spherically
symmetric power-law mass profile allowing for its evolution with
redshift, or iii) total and luminous mass of the lens following
different power-law profiles. We show that the assumed lens model
has a considerable impact on the cosmic curvature constraint found
to be compatible or marginally compatible with the flat case
(depending on the lens model adopted).

Furthermore, we have also divided the SGL sample into three
sub-samples based on lens velocity dispersion: $\sigma_{ap}\leq 200
\; \rm{km~s^{-1}}$ (low-mass galaxies), $200 \;
\rm{km~s^{-1}}<\sigma_{ap}\leq 300 \; \rm{km~s^{-1}}$
(intermediate-mass galaxies) and $\sigma_{ap}>300 \; \rm{km~s^{-1}}$
(high-mass galaxies), and studied the effect on the measurement of
curvature caused by such division. These three lens sub-samples
respectively cover the redshift range of $0.075\leq z_l \leq 0.682$,
$0.082\leq z_l \leq 0.938$, and $0.164\leq z_l \leq 1.004$. Analysis
of low, intermediate and high-mass sub-samples demonstrated that
their best fitted power-law profiles of lens mass density
distribution differ between them. However, they are still consistent
with each other within the uncertainty limits. Discrepancies
regarding the cosmic curvature are less pronounced in this case. One
should bear in mind that it may not be reasonable to characterize
all lenses with a uniform model. More interestingly, we find the SIS
model ($f_E=1$) is well consistent with results obtained in both
low-mass and intermediate-mass (low and intermediate velocity
dispersion) galaxies. Moreover, in the framework of spherically
symmetric power-law mass model, the ranges of slope parameters
($\gamma_0, \gamma_1$) for relatively high-mass galaxies are very
close to the estimates obtained for low-mass and intermediate-mass
elliptical galaxies. Therefore, our analysis indicates the mass
distribution in high-mass lensing galaxies can be effectively
described by the power-law spherical model. More importantly, all of
the full sample and three sub-samples of galaxies, in the framework
of power-law spherical lens model will provide consistent fitting
results, i.e., a closed universe is favored by the current
observational SGL data. Finally, when allowing the luminosity
density profile to be different from the total mass density profile,
high-mass lenses provide clearly different constraints on the slope
parameters ($\alpha, \delta$) compared to their low-mass
counterparts. Such finding in the extended power-law lens model,
which reveals the possible differences in mass density distributions
of dark matter and luminous baryons in massive early-type galaxies,
could help us improve our understanding of the baryonic and dark
matter physics at kpc scale.

Given the limitations of the current sample, we have also performed
a detailed quantitative assessment of what kind of strong lensing
and quasar data are necessary to provide the constraint on cosmic
curvature competitive with the CMBR data. On the one hand, ongoing
and future massive surveys like the LSST will provide up to three
orders of magnitude more galactic-scale strong lenses that are
currently known. On the other hand, the fractional uncertainty of
the angular size of the compact structure in radio quasars will be
significantly reduced by both current and future radio astronomical
surveys. Based on the mock catalogs of strong gravitational lenses
and quasars with the current measurement accuracy, we find that the
forthcoming LSST survey supplemented with the much richer data
regarding ILQSO would be able to constrain cosmic curvature with the
precision comparable to current CMBR data. This encourages us to
consider the possibility of testing the cosmic curvature at the much
higher accuracy with future surveys of strong lensing systems and
high-quality radio astronomical observations of quasars.

\section*{Acknowledgments}
We would like to thank Zheng-Xiang Li and Jun-Qing Xia for their
helpful discussions. This work was supported by National Key R\&D
Program of China No. 2017YFA0402600; the Ministry of Science and
Technology National Basic Science Program (Project 973) under Grants
No. 2014CB845806; the National Natural Science Foundation of China
under Grants Nos. 11503001, 11373014, and 11690023; Beijing Talents
Fund of Organization Department of Beijing Municipal Committee of
the CPC; the Fundamental Research Funds for the Central Universities
and Scientific Research Foundation of Beijing Normal University; and
the Opening Project of Key Laboratory of Computational Astrophysics,
National Astronomical Observatories, Chinese Academy of Sciences.
J.-Z. Qi was supported by the China Postdoctoral Science Foundation
(Grant No. 2017M620661). M. Biesiada was supported by the Foreign Talent
Introduction Project and the Special Fund Supporting Introduction of
Foreign Knowledge Project in China.

\bibliographystyle{mnras}
\bibliography{Okbib}

\bsp    % typesetting comment
\label{lastpage}
\end{document}